\documentclass[aps,prl,twocolumn,superscriptaddress,showpacs]{revtex4}
\usepackage{graphicx}
\usepackage{mathrsfs}
\usepackage{bm}
\usepackage{amsmath}
\usepackage{amsfonts}
\usepackage{multirow}
\usepackage{bigstrut}
\usepackage{rotating}
\usepackage{booktabs}
\usepackage{color}
\usepackage{braket}
\usepackage{dcolumn}
\usepackage{enumerate}
\usepackage{hyperref}
\begin{document}

\title{The General Principle behind Magnetization-induced Second-Order Topological Corner States in the Kane-Mele Model}

\author{Cheng-Ming Miao}
\affiliation{College of Physics, Hebei Normal University, Shijiazhuang 050024, China}

\author{Lizhou Liu}
\affiliation{College of Physics, Hebei Normal University, Shijiazhuang 050024, China}

\author{Yu-Hao Wan}
\affiliation{International Center for Quantum Materials, School of Physics, Peking University, Beijing 100871, China}

\author{Qing-Feng Sun}
\affiliation{International Center for Quantum Materials, School of Physics, Peking University, Beijing 100871, China}
\affiliation{Hefei National Laboratory, Hefei 230088, China}

\author{Ying-Tao Zhang}
\email[]{zhangyt@mail.hebtu.edu.cn}
\affiliation{College of Physics, Hebei Normal University, Shijiazhuang 050024, China}

\begin{abstract}
We propose a general principle for realizing second-order topological corner states in the modified Kane-Mele model with magnetization. It is demonstrated that the sign of the edge Dirac mass depends on the magnetization of the edge sublattice termination. By adjusting the directions of magnetization according to the type of sublattice at the termination of two edges, a mass domain wall can be induced in the presence of topological corner states at an arbitrary position. All previous work on introducing magnetization in the Kane-Mele model to realize second-order topological corner states can be explained by the presence of the Dirac mass domain wall with opposite signs. Applying this principle, we design square-shaped and armchair-type hexagon-shaped graphene nanoflakes with edge magnetization, allowing for the emergence of second-order topological corner states. Our findings serve as a general theory, demonstrating that the realization of second-order topological corner states is not limited by boundary type or nanoflake shape.
\end{abstract}
\maketitle
\section{\uppercase\expandafter{\romannumeral 1}. Introduction}

A two-dimensional topological insulator,
also referred to as a quantum spin Hall insulator,
is capable of supporting spin-helical gapless edge states protected
by time-reversal symmetry \cite{Kane2005,Kane2005a,Bernevig2006,Hasan2010,Moore2010,Qi2011,Bansil2016,Chiu2016,Ren2016}.
The breaking of time-reversal symmetry and the opening of an edge band gap in a topological insulator are essential steps towards observing other quantum states.
One such state discovered is a higher-order topological state \cite{Benalcazar2017,Benalcazar2017a,Song2017,Langbehn2017,Peng2017,Ezawa2018,Ezawa2018a,Schindler2018,Kunst2018,Khalaf2018,Fukui2018,Park2019,Banerjee2020}, where the insulating $d$-dimensional bulk hosts ($d-n$)-dimensional edge states with $n>1$.
It is well known that exploiting the magnetic proximity effect and
conventional transition metal doping are two general approaches to breaking time-reversal symmetry \cite{Larson2008,Wei2013,Vergniory2014}.
Based on this, various magnetic orderings, including in-plane ferromagnetism, antiferromagnetism, and alternate magnetism, are introduced into two-dimensional topological systems to realize second-order topological phases which host topological in-gap zero-dimensional corner states \cite{Sheng2019,Ren2020,Zeng2022,Han2022,Zhu2022,Miao2022,Miao2023,Zhu2023,Zhu2023a}.
The origin of these topological corner states can be readily understood from the perspective of Dirac equations: the two original helical edge states of the first-order topological insulator are gapped by Dirac mass of opposite sign, leading to the formation of a Dirac mass domain wall at the corner. According to the Jackiw-Rebbi theory \cite{Jackiw1976}, such a Dirac mass domain wall would bind a zero-dimensional bound state, namely a second-order topological corner state \cite{Ezawa2018b,Yan2019,Ren2020,Ghosh2020,Chen2020,Yang2020,Zhuang2022}.

Since the Kane-Mele model serves as an ideal platform for realizing the quantum spin Hall effect, several works on this model have demonstrated the occurrence of second-order topological corner states through the application of an in-plane magnetic field or edge magnetic ordering to open the gap of helical edge states \cite{Ren2020,Han2022,Miao2022}.
However, the presence of zigzag edges in graphene nanoflakes is a necessary condition in these proposals, the reason is that the energy gap of the edge states at the armchair boundary is not opened in the presence of in-plane ferromagnetic ordering \cite{Miao2023,Zhu2023}.
To overcome the limitations imposed by boundary types, Zhu $et~al$. have predicted the emergence of topological corner states at zigzag-beard edge domain walls \cite{Zhu2022, Zhu2023}.
Nevertheless, these proposals still heavily rely on the choice of boundaries.
To achieve lattice-insensitive second-order topological corner states, our previous work involved designing a heterojunction comprising two square-shaped graphene nanoflakes with opposite in-plane antiferromagnetic orderings \cite{Miao2023}.
This configuration induces topologically protected zero-dimensional in-gap states. However, it was not possible to realize topological corner states in a single square-shaped nanoflake. Consequently, the exact relationship between magnetization and the sign of the Dirac mass term at the boundary remains unclear, limiting the formation of second-order topological corner states by boundary type and nanoflake shape.

\begin{figure}
	\centering
	\includegraphics[width=1\columnwidth,clip]{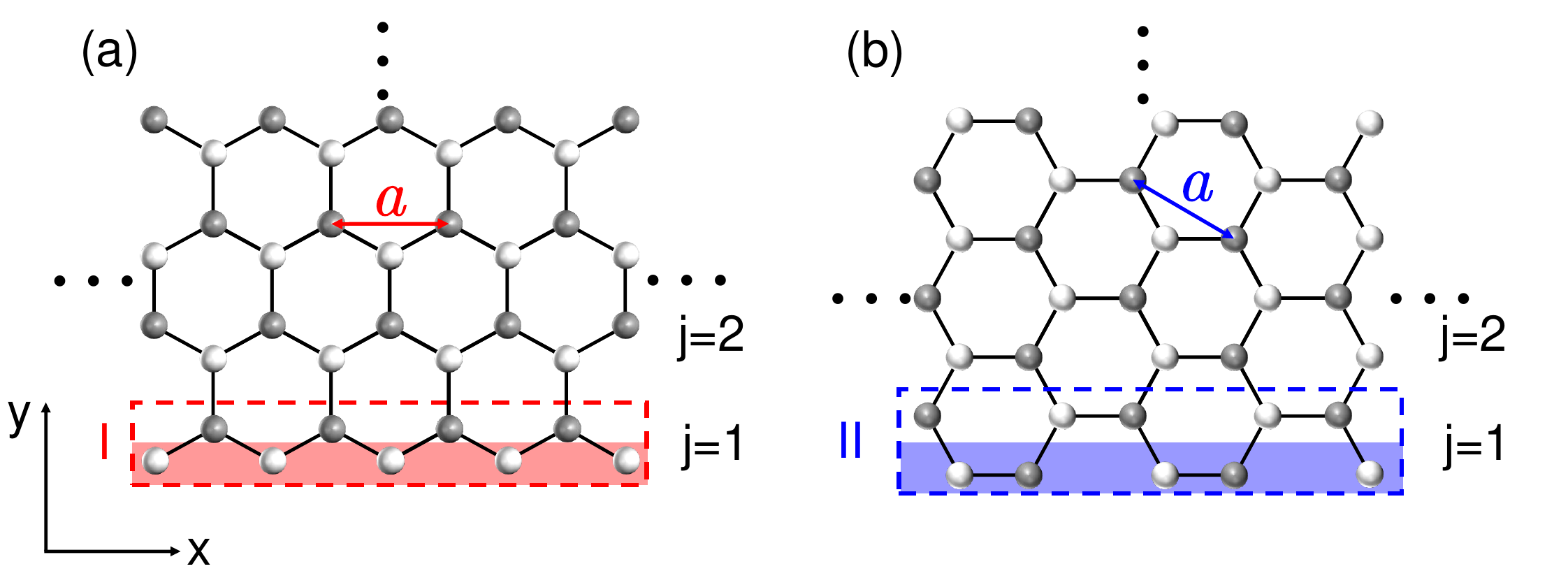}
	\caption{The semi-infinite honeycomb lattice of a lattice constant $a$ with zigzag edge for (a) and armchair edge for (b). Sublattices $A$ and $B$ are denoted by white and grey balls, respectively.
	The choice of super unit cells for the zigzag edge and the armchair edge are illustrated by the red dashed boxs and the blue dashed boxs, respectively.
	The corresponding colored square circled sublattices are the edge sublattice termination of the half-infinite sample area $y>0$.
	Here, $j$ is a real-space row index in $y$ direction perpendicular to the edge.
	}
	\label{fig1}
\end{figure}

In this paper, we present an analytic approach to understanding the second-order topological corner states of the modified Kane-Mele model with magnetization.
Our analysis reveals that the presence of Dirac mass domain walls is a necessary condition for second-order topological corner states. Furthermore, we find that the sign of the edge Dirac mass depends on the magnetization of the edge sublattice termination.
To achieve a non-zero Dirac mass term at the armchair boundary, it is necessary to establish in-plane antiferromagnetic orderings with $AB$ sublattices oriented in opposite directions. The sign of the Dirac mass term corresponds to the magnetization direction at sublattice $A$.
At the zigzag edge, where the outermost sublattice is $A$ ($B$), the Dirac mass sign aligns with the same (opposite) magnetization direction at sublattice $A$ ($B$).
This general edge magnetic principle can explain all previous work on introducing magnetization in the Kane-Mele model to obtain topological corner states.
Applying this principle, we design both square-shaped nanoflakes with zigzag and armchair edges and hexagon-shaped nanoflakes with armchair edges in the presence of edge magnetization, where second-order topological corner states can be obtained. This demonstrates that the realization of second-order topological corner states is not limited by boundary type and nanoflake shape, providing a general theory.

The rest of the paper is organized as follows. In Sec. II, we give general edge magnetic principle to obtain second-order topological corner states in the modified Kane-Mele model.
In Sec. III, we apply general edge magnetic principle to design the square-shaped and hexagon-shaped graphene nanoflakes with edge magnetization to realize second-order topological corner states. Finally, a summary is presented in Sec. IV.

\section{\uppercase\expandafter{\romannumeral 2}. General principle of the corner states in the Kane-Mele model with magnetization}

In this section, we derive the self-consistent equations for the edge state of the Kane-Mele model in the semi-infinite lattice, as shown in Fig. \ref{fig1}.
We start from the Kane-Mele model \cite{Kane2005}, which is the tight-binding model with a spin-orbit interaction in a honeycomb lattice:
\begin{align}
	H_{KM}=-t\sum_{\left< i,j \right> ,s}{c_{is}^{\dag}c_{js}}+\sum_{\left< \left< i,j \right> \right> ,s ,s'}{i t_{s} \nu _{ij} c_{is}^{\dag}c_{js'}\left[ \mathbf{\hat{s}}_z \right] _{s s'}},
	\label{eq1}
\end{align}
where $c_{is}$ and $c_{is}^{\dag}$ are the annihilation and creation operators for an electron at the discrete site $i$ with spin $s=\uparrow(\downarrow)$.
The first term represents the nearest-neighbor hopping, and the second term corresponds to the intrinsic spin-orbit interaction with an amplitude of $t_{s}$, connecting the second nearest neighbors. Here, $\nu_{ij}=+1$ ($-1$) denotes that the electron makes a left (right) turn during hopping from site $j$ to $i$.

As depicted in Fig. \ref{fig1}(a), we consider a half-infinite sample area where $y>0$ with a zigzag boundary in the $x$-direction.
The tight-binding Hamiltonian can be expressed in momentum $q$ in the $x$-direction and in the real space lattice index $j$ in the $y$-direction as \cite{Tan2022}:
\begin{align}
	&H(q)=\sum_{j}{\hat{\Psi}_{j}^{\dagger}(q) T_{0}^{Z}(q)\hat{\Psi}_{j}(q)
	 +\hat{\Psi}_{j-1}^{\dagger}(q) T_{+1}^{Z}(q)\hat{\Psi}_{j}(q)}\nonumber \\
	&\ \ \ \  \ \ \ \ +\hat{\Psi}_{j+1}^{\dagger}(q) T_{-1}^{Z}(q)\hat{\Psi}_{j}(q),\nonumber \\
	&T_{0}^{Z}(q)=2 t_{s} \sin {q} s_z \sigma_z+2 t \cos \frac{q}{2} s_0 \sigma_x,\nonumber \\ &T_{+1}^{Z}(q)=-2 t_{s} \sin \frac{q}{2} s_z \sigma_z+t s_0 \left(\frac{\sigma_x-i \sigma_y}{2}\right),\nonumber \\
	&T_{-1}^{Z}(q)=-2 t_{s} \sin \frac{q}{2} s_z  \sigma_z+t s_0 \left(\frac{\sigma_x+i \sigma_y}{2}\right),
	\label{eq2}
\end{align}
where $\sigma _i^{}$ and $s_i^{}$ are Pauli matrices in the sublattice
$\left( {A, B} \right)$ and spin $\left( { \uparrow , \downarrow } \right)$. The basis vector is chosen as $\hat{\Psi}_{j}(q)
=(\hat{\Psi}_{j,A,\uparrow}(q),\hat{\Psi}_{j,B,\uparrow}(q),
\hat{\Psi}_{j,A,\downarrow}(q), \hat{\Psi}_{j,B,\downarrow}(q))^{T} $.
We construct the Harper's equation \cite{Harper1955} which describes the wave function in real space in the direction normal to the edge as:
\begin{eqnarray}
&E(q)\Psi_{j}(q)=  T_{0}^{Z}(q)\Psi_{j}(q)+ T_{+1}^{Z}(q)\Psi_{j+1}(q)\nonumber \\
&+ T_{-1}^{Z}(q)\Psi_{j-1}(q),
\label{eq3}
\end{eqnarray}
with the wave function $\Psi_{j}(q)$.
The $j$ dependence of the wave function can be simplified by setting $\Psi_{j}(q)=e^{\kappa(j-1)}\Psi_{1}(q)$.
So the relation between the wave function of the $j+1$ ($j-1$) row and the $j$ row is given by $\Psi_{j+1}(q)=e^{\kappa}\Psi_{j}(q)$ ($\Psi_{j-1}(q)=e^{-\kappa}\Psi_{j}(q)$).
We can rewrite Eq. (\ref{eq3}) as follow:
\begin{eqnarray}
E(q)\Psi_{j}(q)&=& [T_{0}^{Z}(q)+ T_{+1}^{Z}(q)e^{\kappa}+ T_{-1}^{Z}(q)e^{-\kappa}] \Psi_{j}(q)\nonumber \\
    &=&\mathcal{H}^{I}_{Z}\Psi_{j}(q),
\label{eq}
\end{eqnarray}
where $\mathcal{H}^{I}_{Z}$ is the effective Hamiltonian of edge I. Utilizing Eqs. (\ref{eq2}) and (\ref{eq}), the effective Hamiltonian of edge I can be written as:
\begin{eqnarray}
	&\mathcal{H}^{I}_{Z}&=T_{0}^{Z}(q)+ e^{\kappa}T_{+1}^{Z}(q)+e^{-\kappa}T_{-1}^{Z}(q) \nonumber \\
	&&=4 t_{s} \sin \frac{q}{2}\left(\cos \frac{q}{2}-\cosh \kappa\right) s_z\sigma_z-i t \sinh \kappa s_0\sigma_y\nonumber \\
	&&+\left(2 t \cos \frac{q}{2}+t \cosh \kappa\right) s_0\sigma_x.
	\label{eq4}
\end{eqnarray}
The helical gapless edge states cross at the time-reversal invariant point $q=\pi$ with energy $E=0$.
In this condition, the eigenstates of the spin-decouped edge Hamiltonian can be represented by the following spinors \cite{Tan2022,Miao2023}:
\begin{eqnarray}
	&&\left|\chi_{\uparrow}^{I} \right>=\frac{1}{\sqrt{1+\mu}} \left[\begin{array}{cc}
			1 \\
			0
		\end{array}\right]_{s} \otimes \left[\begin{array}{cc}
			1 \\
			-i\sqrt{\mu}
		\end{array}\right]_{\sigma},\nonumber \\
	&&\left|\chi_{\downarrow}^{I} \right>=\frac{1}{\sqrt{1+\mu}} \left[\begin{array}{cc}
			0 \\
			1
		\end{array}\right]_{s} \otimes \left[\begin{array}{cc}
			1 \\
			i\sqrt{\mu}
		\end{array}\right]_{\sigma},
	\label{eq5}
\end{eqnarray}
with $\mu=(1+\frac{t^2}{8t_{s}^{2}})-\sqrt{(1+\frac{t^2}{8t_{s}^{2}})^2-1}$. Similarly, the eigenstates of the armchair edge II [see Fig. \ref{fig1}(b)] turn out to be \cite{Tan2022,Miao2023}:
\begin{eqnarray}
	\left|\chi_{\uparrow}^{II} \right>= \left[\begin{array}{cc}
		1 \\
		0
	\end{array}\right]_{s} \otimes \left[\begin{array}{cc}
		1 \\
		-i
	\end{array}\right]_{\sigma}, \left|\chi_{\downarrow}^{II} \right>= \left[\begin{array}{cc}
		0 \\
		1
	\end{array}\right]_{s} \otimes \left[\begin{array}{cc}
		1 \\
		i
	\end{array}\right]_{\sigma}.
	\label{eq6}
\end{eqnarray}

It is noteworthy that for a half-infinite sample area where $y>0$,
the outermost sublattices of the zigzag edge $I$ are all sublattice $A$.
However, for $y<0$, the outermost sublattices of edge $I$ switch to sublattice $B$, as illustrated in Fig. \ref{fig1}(a).
On the other hand, the outermost sublattices of the armchair edge $II$ alternate between sublattices $A$ and $B$, regardless of whether the half-infinite sample area is $y>0$ or $y<0$.
The selection of the half-infinite region does not affect the eigenstates given by Eqs. (\ref{eq5}) and (\ref{eq6}) on edges $I$ and $II$.

In the presence of the in-plane magnetization, the Hamiltonian of the system can be written by the adding the magnetization term to Eq. (\ref{eq1}):
\begin{align}
	H=H_{KM}+J \sum_{i,s,s'}{c_{is}^{\dag}c_{is'}\left[ \mathbf{\hat{n} \cdot \hat{s}} \right] _{s s'}},
	\label{eq7}
\end{align}
where $J=J_{A/B}$ stands for the magnetization strength at sublattice $A/B$. The orientation of magnetization is along the unit vector $\mathbf{\hat{n}}=(n_x, n_y, 0)$.
Without loss of generality, we take $\mathbf{\hat{n}}$
along the $\hat{y}$ direction.
The effective mass term ${\bf M}$ of the edges can be obtained by incorporating the in-plane magnetization term $Js_y$ into the subspace
of the edge states. Additionally, if the magnetization direction is along the $\hat{x}$ direction, the in-plane magnetic field term becomes $Js_x$, which does not affect the calculation of the Dirac mass term.
Since the edge states are highly localized to the outermost sublattices,
we consider adding ferromagnetism to the outermost sublattices A and B, respectively.
The two Dirac mass terms at zigzag edge $I$ can then be expressed as follows:
\begin{eqnarray}
	&{{\bf M}_{I_{A}}}&={J_A}\left( {\begin{array}{*{20}{c}}
			{\langle \chi _ \uparrow ^{I}|\frac{{\sigma _0}+{\sigma _z}}{2}{s_y}\left| {\chi _ \uparrow ^{I}} \right\rangle }&{\langle \chi _ \uparrow ^{I}|\frac{{\sigma _0}+{\sigma _z}}{2}{s_y}\left| {\chi _ \downarrow ^{I}} \right\rangle }\\
			{\langle \chi _ \downarrow ^{I}|\frac{{\sigma _0}+{\sigma _z}}{2}{s_y}\left| {\chi _ \uparrow ^{I}} \right\rangle }&{\langle \chi _ \downarrow ^{I}|\frac{{\sigma _0}+{\sigma _z}}{2}{s_y}\left| {\chi _ \downarrow ^{I}} \right\rangle }
	\end{array}} \right)\nonumber\\
	&&=\frac{J_{A}}{{{\rm{|1 + }}\mu {\rm{|}}}}\left( {\begin{array}{*{20}{c}}
			0&-i\\
			i&0
	\end{array}} \right)  =\frac{{J_A}}{{{\rm{|1 + }}\mu {\rm{|}}}}{s_{y}}
	= m_{I_{A}}^{}{s_{y}},\nonumber \\
	&{{\bf M}_{I_{B}}}&={J_B}\left( {\begin{array}{*{20}{c}}
			{\langle \chi _ \uparrow ^{I}|\frac{{\sigma _0}-{\sigma _z}}{2}{s_y}\left| {\chi _ \uparrow ^{I}} \right\rangle }&{\langle \chi _ \uparrow ^{I}|\frac{{\sigma _0}-{\sigma _z}}{2}{s_y}\left| {\chi _ \downarrow ^{I}} \right\rangle }\\
			{\langle \chi _ \downarrow ^{I}|\frac{{\sigma _0}-{\sigma _z}}{2}{s_y}\left| {\chi _ \uparrow ^{I}} \right\rangle }&{\langle \chi _ \downarrow ^{I}|\frac{{\sigma _0}-{\sigma _z}}{2}{s_y}\left| {\chi _ \downarrow ^{I}} \right\rangle }
	\end{array}} \right)\nonumber\\
	&&=-\frac{{J_B}{{\rm{|}}\mu {\rm{|}}}}{{{\rm{|1 + }}\mu {\rm{|}}}}\left( {\begin{array}{*{20}{c}}
			0&-i\\
			i&0
	\end{array}} \right)  =-\frac{{J_B}{{\rm{|}}\mu {\rm{|}}}}{{{\rm{|1 + }}\mu {\rm{|}}}}{s_{y}}
	= m_{I_{B}}^{}{s_{y}}.
	\label{eq8}
\end{eqnarray}

It is widely acknowledged that a Dirac mass domain wall forms
when the Dirac mass terms have opposite signs. Thus, adding magnetism in the same direction to sublattices $A$ and $B$ necessarily results in Dirac mass domain walls, as indicated by $m_{I_{A}} m_{I_{B}}<0$ according to Eq. (\ref{eq8}). This explains why the corner states in Ref. \cite{Ren2020} appear at the two obtuse corners of the diamond-shaped nanoflake with zigzag boundaries when introducing in-plane ferromagnetic orderings.
The rationale is that the outermost sublattices on either side of the obtuse corner are sublattices $A$ and $B$, respectively. Similarly, all previous works \cite{Han2022,Miao2022,Miao2023,Zhu2023} on second-order corner states in zigzag-boundary or beard-boundary graphene nanoflakes can be explained by the presence of Dirac mass domain walls with opposite signs. The signs of the Dirac mass terms all satisfy Eq. (\ref{eq8}), which demonstrates that the edge magnetic principle we obtained is universal.

However, there are both sublattices $A$ and $B$ at the outermost layer of the armchair edge $II$, so both sublattices $A$ and $B$ need to be introduced magnetization.
With in-plane ferromagnetic ordering, the armchair edges maintain gapless helical edge states \cite{Miao2023}.
Conversely, the gapless edge states on the armchair edges can be gapped by in-plane antiferromagnetic ordering \cite{Miao2023, Zhu2023}.
To find out the reason for the difference of edge states between ferromagnetism and antiferromagnetism, we derive the effective mass term ${\bf M}_{II}^{FM}$ (${\bf M}_{II}^{AFM}$) of the armchair edge $II$ by projecting the in-plane ferromagnetic (antiferromagnetic) ordering term $J_A\sigma_0s_y$ ($J_A\sigma_zs_y$) onto the spin subspace spanned by $\left| {\chi _ \uparrow ^{II}} \right\rangle$ and $\left| {\chi _ \downarrow ^{II}} \right\rangle$.
\begin{eqnarray}
	 &{{\bf M}_{II}^{FM}}&= {J_A}\left( {\begin{array}{*{20}{c}}
			{\langle \chi _ \uparrow ^{II}|{\sigma _0}{s_y}\left| {\chi _ \uparrow ^{II}} \right\rangle }&{\langle \chi _ \uparrow ^{II}|{\sigma _0}{s_y}\left| {\chi _ \downarrow ^{II}} \right\rangle }\\
			{\langle \chi _ \downarrow ^{II}|{\sigma _0}{s_y}\left| {\chi _ \uparrow ^{II}} \right\rangle }&{\langle \chi _ \downarrow ^{II}|{\sigma _0}{s_y}\left| {\chi _ \downarrow ^{II}} \right\rangle }
	\end{array}} \right)\nonumber\\
	&&  = {J_A}\left( {\begin{array}{*{20}{c}}
			0&0\\
			0&0
	\end{array}} \right)=0*{s_y}=m_{II}^{FM}s_y,\nonumber \\
&{{\bf M}_{II}^{AFM}}&= {J_A}\left( {\begin{array}{*{20}{c}}
			{\langle \chi _ \uparrow ^{II}|{\sigma _z}{s_y}\left| {\chi _ \uparrow ^{II}} \right\rangle }&{\langle \chi _ \uparrow ^{II}|{\sigma _z}{s_y}\left| {\chi _ \downarrow ^{II}} \right\rangle }\\
			{\langle \chi _ \downarrow ^{II}|{\sigma _z}{s_y}\left| {\chi _ \uparrow ^{II}} \right\rangle }&{\langle \chi _ \downarrow ^{II}|{\sigma _z}{s_y}\left| {\chi _ \downarrow ^{II}} \right\rangle }
	\end{array}} \right)\nonumber\\
	&&  = 2{J_A}\left( {\begin{array}{*{20}{c}}
			0&-i\\
			i&0
	\end{array}} \right)=2{J_A}{s_y}=m_{II}^{AFM}s_y.
\label{eq9}
\end{eqnarray}

From Eq. (\ref{eq9}), it is evident that introducing ferromagnetic ordering $J_A\sigma_0s_y$ at armchair edge $II$ yields a zero mass term, thereby preserving gapless helical edge states.
Conversely, with antiferromagnetic ordering, the Dirac mass terms
at the armchair edge become non-zero and align in the same direction as the magnetization at sublattice $A$.
Additionally, in the case where the magnetic sizes of sublattices $A$ and $B$ are unequal, it can be decomposed into a linear combination of antiferromagnetic and ferromagnetic terms on sublattices $A$ and $B$.
In this scenario, we can extract the antiferromagnetic component,
which determines the sign of the mass term at the armchair boundary.

To manipulate the sign of the Dirac mass at armchair edge $II$, one can consistently set the magnetization of sublattice $B$ to be opposite that of sublattice $A$, and then adjust the magnetization at sublattice $A$ to be positive or negative \cite{Miao2023}. This method enables the flexible alteration of the Dirac mass sign at armchair edge $II$.


It is evident that achieving a zero-dimensional second-order topological corner state from a two-dimensional first-order topological state involves two steps, as per the edge state theory \cite{Tan2022}.
Firstly, it is imperative to open the energy gap of the first-order topological edge states by breaking the time-reversal symmetry protecting the edge states. Both ferromagnetic and antiferromagnetic orderings can accomplish this in the Kane-Mele model.
However, at the armchair boundary, the presence of ferromagnetism maintains gapless edge states, necessitating the introduction of antiferromagnetism to open the energy gap.
Conversely, at the zigzag edge, the edge states can be gapped by either ferromagnetism or antiferromagnetism.
Secondly, for the corner to be composed of two gapped edge states, the Dirac mass terms of these states must have opposite signs.
Given that antiferromagnetism is introduced at the armchair edge, the Dirac mass term at the armchair edge aligns with the magnetic ordering at sublattice $A$, necessitating it to be opposite in sign to the magnetic ordering at sublattice $B$.
It's noteworthy that for zigzag boundaries, the outermost sublattice type plays a crucial role: if the outermost sublattice is $A$, the sign of the Dirac mass term aligns with its magnetic ordering; conversely, if the outermost sublattice is $B$, the sign of the mass term opposes its magnetic ordering.
Additionally, the magnitude and direction of the magnetic ordering in the second outermost layer do not influence the sign of the Dirac mass term at zigzag edges.
By adjusting the directions of magnetization according to the type of sublattice at the termination of two edges, a mass domain wall can be induced in the presence of topological corner states at an arbitrary position. Consequently, we can set the direction of magnetization of the outermost sublattice according to the different edge states to realize the Dirac mass domain wall with topological corner states. This approach is not limited by boundary type or nanoflake shape.

\section{\uppercase\expandafter{\romannumeral 3}. Application of the genernal principle}
\begin{figure}
	\centering
	\includegraphics[width=1\columnwidth,clip]{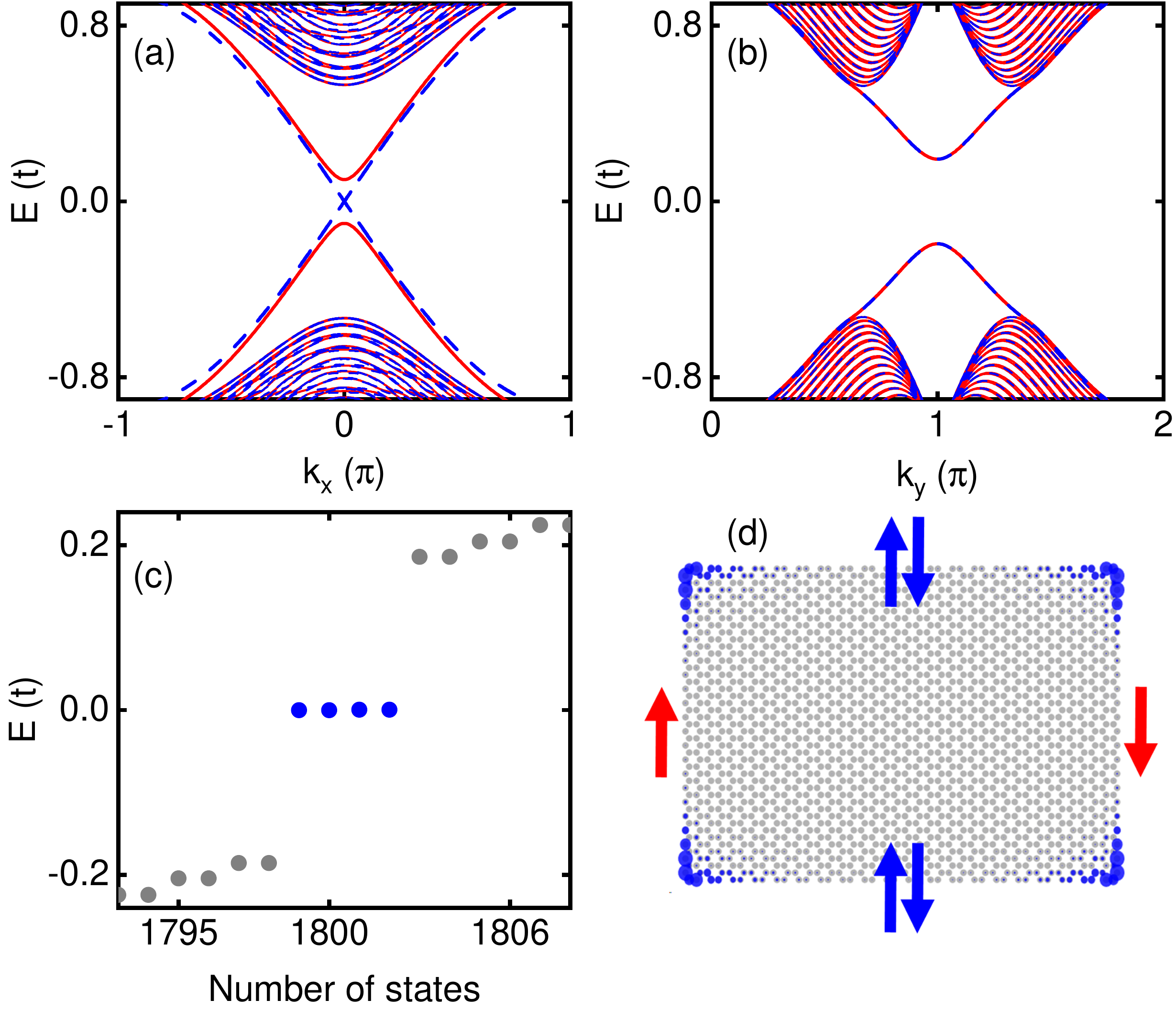}
	\caption{ Band structure of the nanoribbons with edge ferromagnetic ordering (blue dotted lines) and edge antiferromagnetic ordering (red solid lines) for armchair edge in (a) and zigzag edge in (b). (c) Energy levels of square-shaped nanoflake in (d) with the outermost edge magnetization. The blue and grey dots represent the in-gap corner states and gapped edge states, respectively.
	(d) Probability distribution of the in-gap states, where the area of the blue circles are proportional to the charge. The magnetization directions at the four edge sublattice termination are indicated by red and blue arrows. The edges with the same colored magnetization arrows have the same sign of the Dirac mass term. Chosen parameters are  $t_{s}=0.1t,~J=0.2t,~J_{A}^{II}=J_{A}^{IV}=J_{B}^{I}=J,~J_{B}^{II}=J_{B}^{IV}=J_{A}^{III}=-J$.
	}
	\label{fig2}
\end{figure}

We employ the general edge magnetic principle on a square-shaped graphene nanoflake with the outermost edge magnetization, where the armchair (zigzag) edge aligns with the $x$ ($y$) axis. The edge magnetism in graphene nanoflakes can be effectively manipulated \cite{Miao2022,Zhao2024}.
Edge ferromagnetism (antiferromagnetism) indicates that sublattices $A$ and $B$ at the edge sublattice termination share the same (opposite) magnetic orderings, denoted as $J_A=J_B=J$ ($J_A=-J_B=J$).
In Fig.~\ref{fig2}(a), we calculate the band energy structure of the graphene armchair nanoribbon with edge ferromagnetism and antiferromagnetism.
It is demonstrated that a pair of spin-helical gapless edge modes still counterpropagate along an armchair Kane-Mele nanoribbon with edge ferromagnetic orderings [see blue dotted lines in Fig.~\ref{fig2}(a)].
In the presence of the edge antiferromagnetic orderings, the spin-helical gapless edge states are gapped as illustrated by red solid lines in Fig.~\ref{fig2}(a).
We also calculate the band energy structure of the graphene zigzag nanoribbon with edge ferromagnetism and antiferromagnetism in Fig.~\ref{fig2}(b).
Figure~\ref{fig2}(b) shows that the spin-helical gapless edge states are gapped by introducting either edge ferromagnetism or edge antiferromagnetism.
The results of the energy band can be well explained by Eqs. (\ref{eq8}) and (\ref{eq9}): only the ferromagnetic mass term of the armchair boundary is zero corresponding to the gapless edge states, while the other three cases are all non-zero corresponding to the gapped edge states.

Moreover, we plot the energy levels of square-shaped nanoflake with outermost edge magnetization in Fig.~\ref{fig2}(c).
The magnetic ordering directions of the outermost sublattice $B$ ($A$) of the left (right) zigzag edge is along the $+y$ $(-y)$ axis, as illustrated by red arrows in Fig.~\ref{fig2}(d).
Both the top and bottom armchair edges exhibit edge antiferromagnetic ordering along the $+y$ $(-y)$ axis at the sublattice $A(B)$ [see blue arrows in Fig.~\ref{fig2}(d)].
We label the left, top, right and bottom edges of a square as $I$, $II$, $III$ and $IV$, respectively.
One can see that four zero-energy in-gap states emerge, as displayed by the blue dots in Fig.~\ref{fig2}(c).
The probability distribution of a wave function at half-filling is highlighted in Fig.~\ref{fig2}(d). It can be seen that the corner states with fractional charge $e/2$ are almost localized at the four corners of the square-shaped nanoflake.
According to Eqs. (\ref{eq8}) and (\ref{eq9}), the signs of the Dirac mass at four boundaries are $m_{I}<0, m_{III}<0, m_{II}=m_{IV}>0$. This magnetization configuration leads to the formation of four Dirac mass domain walls, each encapsulating a corner state.
Our results fully demonstrate that the magnetization of the outermost sublattice is fundamental to the emergence of the Dirac mass domain wall, which bounds the topological in-gap corner states. The realization of the domain wall is not limited by the boundary type and nanoflake shape, which verifies the general edge magnetic principle.

 \begin{figure}
	\centering
	\includegraphics[width=1\columnwidth,clip]{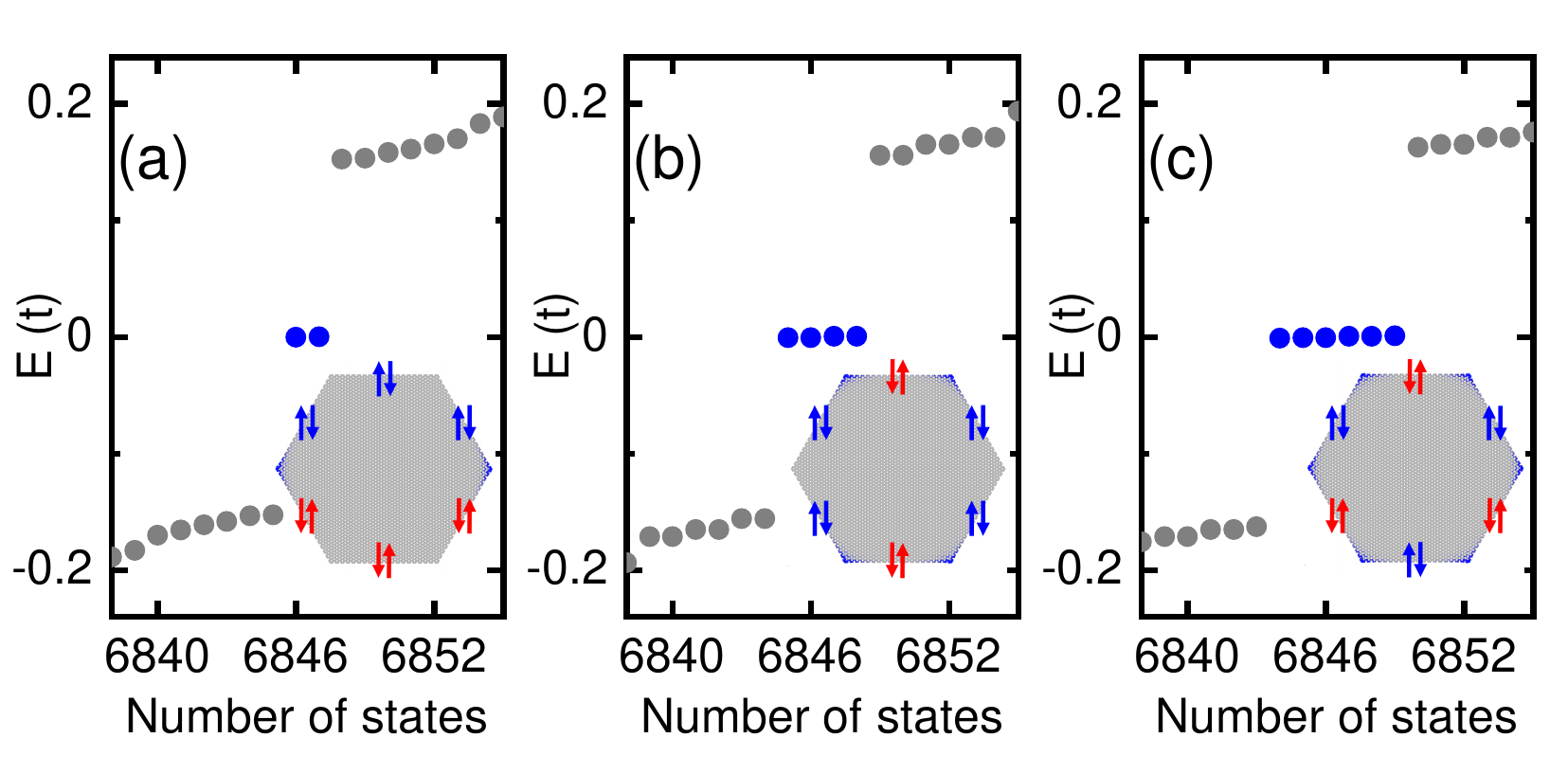}
	\caption{ Energy levels of hexagon-shaped nanoflake with different armchair edge antiferromagnetic orderings $J_{A}^{I}=J_{A}^{II}=J_{A}^{III}=J$, $J_{A}^{IV}=J_{A}^{V}=J_{A}^{VI}=-J$ for (a), $J_{A}^{I}=J_{A}^{III}=J_{A}^{IV}=J_{A}^{VI}=J$, $J_{A}^{II}=J_{A}^{V}=-J$ for (b) and $J_{A}^{I}=J_{A}^{III}=J_{A}^{V}=J$, $J_{A}^{II}=J_{A}^{IV}=J_{A}^{VI}=-J$ for (c). The blue dots represent the in-gap corner states. Probability distribution of the corner states are plotted in the inset.
	The area of the blue circles are proportional to the charge.  The magnetization directions at the four edge sublattice termination are indicated by red and blue arrows. The edges with the same colored magnetization arrows have the same sign of the Dirac mass term. The other parameters are set to be $t_{s}=0.1t,~J=0.3t,~J_{B}=-J_{A}$.
	}
	\label{fig3}
\end{figure}

To demonstrate that the generation of corner states is not limited by the armchair edges, we design a hexagon-shaped graphene nanoflake consisting of six armchair edges. The upper-left edge is labeled as $I$, and the remaining five edges in the clockwise direction are labeled as $II$ to $VI$, respectively.
In Figs. \ref{fig3}(a)-\ref{fig3}(c), we plot the energy levels of the hexagon-shaped nanoflake with different armchair edge antiferromagnetic orderings $J_{B}=-J_{A}$. There are two zero-energy in-gap states for $+y$ direction magnetization at sublattice $A$ of the armchair edges $I/II/III$ (blue arrows) and $-y$ direction magnetization at aublattice $A$ of the armchair edges $IV/V/VI$ (red arrows), as displayed by the blue dots in Fig. \ref{fig3}(a).
The probability distribution of a wave function is highlighted in the inset of Fig. \ref{fig3}(a). It can be seen that the zero-energy in-gap states are almost locatized at the domain wall between blue and red arrows.
If the directions of antiferromagnetic orderings at the sublattice $A$ of the edge $I/III/IV/VI$ ($II/V$) are turn to be $+y$ ($-y$) axis, four zero-energy in-gap states appear [see Fig. \ref{fig3}(b)]. The inset of Fig. \ref{fig3}(b) shows that the four zero-energy in-gap states are distributed at the four corner positions of the hexagon-shaped graphene nanoflake.
In Fig. \ref{fig3}(c), six zero-energy in-gap states (in blue) arise with alternating edge antiferromagnetic orderings, and the wave function distributions of these states (blue density dots) are highlighted in the inset of Fig. \ref{fig3}(c). These six zero-energy in-gap states are bound to six corner positions of the hexagon-shaped graphene nanoflake.
Our results fully demonstrate that the armchair-type edge is not a limiting factor for the emergence of corner states. Due to the Dirac mass domain wall been freely inducible, the number and position of corner states can be arbitrarily adjusted in the armchair-type hexagon-shaped graphene nanoflake, as long as it satisfies the general edge magnetic principle.
Furthermore, in magnetized graphene nanoflakes of any shape and boundary,
one can achieve second-order topological corner states that are customizable in both number and position. It only requires setting the magnetic orderings of the two edges of the component corner according to the principle of the opposite sign of the Dirac mass term. Thus our findings are universal, independent of boundary type and nanoflake shape.

\section{\uppercase\expandafter{\romannumeral 4}. summary}

In summary, we propose an analytic approach to understanding second-order topological corner states in the modified Kane-Mele model with magnetization.
It indicates that the existence of second-order topological corner states is contingent upon the presence of Dirac mass domain walls with opposite signs.
The sign of the Dirac mass term depends on the direction of the magnetic ordering at the outermost sublattice.
Specifically, the armchair edge exhibits alternating $A$ and $B$ sublattices in its outermost layer, while the zigzag edge features either all $A$ or all $B$ sublattices in its outermost layer.
The introduction of ferromagnetic ordering at armchair edge results in a zero mass term, whereas armchair edges with antiferromagnetic ordering have non-zero Dirac mass terms with the same direction as the magnetization at sublattice $A$.
At the zigzag edge, where the outermost sublattice is $A$ ($B$), the Dirac mass sign aligns with the same (opposite) magnetization direction at sublattice $A$ ($B$).
This general edge magnetic principle elucidates all previous work on introducing magnetism to the Kane-Mele model for obtaining topological corner states.
Applying this principle, we study the energy band and energy levels of square-shaped and hexagon-shaped graphene nanoflakes with edge magnetism.
It is found that the magnetization of the outermost sublattice is fundamental to the emergence of the Dirac mass domain wall, which bounds the topological in-gap corner states.
Futhermore, in graphene nanoflakes of any shape and boundary, the number and position of corner states can be freely adjusted by setting the magnetic orderings of the two edges of the component corner, adhering to the principle of the opposite sign of the Dirac mass term.
It demonstrates that the realization of the second-order topological corner states is not limited by boundary type and nanoflake shape, emphasizing the universality of our findings.

\begin{acknowledgments}
\section{acknowledgments}
This work was financially supported by the National Natural Science Foundation of China (Grant No. 12074097, No. 12374034, and No. 11921005),
Natural Science Foundation of Hebei Province (Grant No. A2020205013),
the Innovation Program for Quantum Science and Technology (2021ZD0302403),
and the Strategic Priority Research Program of Chinese Academy of Sciences (Grant No. XDB28000000).
\end{acknowledgments}


\end{document}